\documentclass[aps,showpacs,floats,onecolumn,superscriptaddress,longbibliography,preprint]{revtex4-1}
\usepackage[utf8]{inputenc}
\usepackage[english]{babel}
\usepackage[]{caption}
\usepackage{amsmath,amssymb,fancyhdr,graphicx,grffile}
\usepackage{subfigure,upgreek}

\newcommand{\eqnref}[1]{Eq.~(\ref{eq:#1})}
\newcommand{\eqanref}[1]{Equation~(\ref{eq:#1})}

\usepackage{color}
\definecolor{tuered}{RGB}{214,0,74}
\definecolor{tueblue}{RGB}{0,102,204}
\definecolor{grey}{RGB}{128,128,128}
\definecolor{specialgreen}{RGB}{0,214,0}

\usepackage{tikz,pgfplots}
\usepackage{relsize}
\tikzset{fontscale/.style = {font=\relsize{#1}}}
\usetikzlibrary{calc}

\begin{document}
\newcommand{\temp}{T}

\newcommand{\kB}{k_B}

\newcommand{\HB}{H}

\newcommand{\shcv}{c_v}

\newcommand{\svol}{V}

\newcommand{\sysi}{L_S}

\newcommand{\pres}{P}

\newcommand{\mfp}{\lambda_f}

\newcommand{\tlsp}{\l_f}

\newcommand{\entr}{{\cal S}}

\newcommand{\nflI}{N_1}

\newcommand{\nflII}{N_2}

\newcommand{\nflJ}{N_j}

\newcommand{\nfli}{N_i}

\newcommand{\npemu}{\check{N}_{\rm p}}

\newcommand{\npens}{N_p}

\newcommand{\ndir}{N_i}

\newcommand{\pc}{\Delta\rho}

\newcommand{\pca}{\theta}

\newcommand{\pcaO}{\pca_0}

\newcommand{\pcaK}{\pca_\lt}

\newcommand{\COLfita}{a_{\pca}}
\newcommand{\COLfitb}{b_{\pca}}

\newcommand{\pcaa}{\pca_A}

\newcommand{\pcap}{\pca_P}

\newcommand{\pcapt}{\pca_{P,t}}

\newcommand{\psjp}{\Delta\pca}

\newcommand{\psjpt}{\psjp_t}

\newcommand{\psjptfull}{\psjp_T}

\newcommand{\psjptlinf}{\psjp_{t,\infty}}

\newcommand{\psjpsup}{\psjp_{\rm sup}}

\newcommand{\mmjp}{Mm}

\newcommand{\jb}{J}

\newcommand{\jd}{x_{\rm Janus}}

\newcommand{\jdn}{n_{\rm Janus}}

\newcommand{\ja}{\alpha}

\newcommand{\pr}{R}

\newcommand{\prp}{R_{\parallel}}

\newcommand{\pro}{R_{\perp}}

\newcommand{\proeff}{R_{0}}

\newcommand{\proeffrealif}{r_{0}}

\newcommand{\prpeff}{{\cal R}}

\newcommand{\prpeffflat}{{\cal R}}

\newcommand{\prpefffit}{{\cal R}_{\rm fit}}

\newcommand{\expprp}{a_{\parallel}}

\newcommand{\exppro}{a_{\perp}}

\newcommand{\pvI}{\mathbf{u}_i}

\newcommand{\pvII}{\mathbf{u}_j}

\newcommand{\pvel}{\pv}

\newcommand{\radint}{r}

\newcommand{\radintcuttoff}{R_{\rm max}}

\newcommand{\radintcuttoffO}{{\cal R}_{\rm max}}

\newcommand{\prmax}{R_{\rm max}}

\newcommand{\distpc}{r}

\newcommand{\prnuc}{R_{\rm nuc}}

\newcommand{\rdrop}{R_{\rm drop}}

\newcommand{\rdropO}{R_{\rm drop,0}}

\newcommand{\prnuccrit}{R_{\rm nuc,crit}}

\newcommand{\ucol}{U_{\rm col}}

\newcommand{\ucolif}{U_{\rm col,if}}

\newcommand{\ubulk}{U_{\rm bulk}}

\newcommand{\udirect}{U_{\rm direct}}

\newcommand{\ucap}{U_{\rm cap}}

\newcommand{\ubulkhc}{U_{\rm hc}}

\newcommand{\eci}{\beta}

\newcommand{\Fcapcurved}{F_{\rm cap,curved}}

\newcommand{\Tpc}{D}

\newcommand{\Tpcap}{\Tpc_{\rm cap}}

\newcommand{\TpcapO}{\Tpc_{\rm cap,0}}

\newcommand{\Tpcappoint}{\Tpc_{\rm cap,point}}

\newcommand{\Tpcappointpm}{\Tpc_{\rm cap,point,\pm}}

\newcommand{\dTpcappoint}{\Delta\Tpc_{\rm cap,point}}

\newcommand{\ifprincurv}{\zeta}

\newcommand{\ifprincurvI}{\ifprincurv_1}

\newcommand{\ifprincurvII}{\ifprincurv_2}

\newcommand{\ifprincurvJ}{\ifprincurv_i}

\newcommand{\ifmeanprincurv}{{\cal H}}

\newcommand{\ifgprincurv}{{\cal G}}

\newcommand{\geocurvIIIpcl}{{\cal K}}

\newcommand{\dpres}{\delta\pres}

\newcommand{\spacialx}{\vec{x}}

\newcommand{\potH}{\phi_H}

\newcommand{\constH}{K_H}

\newcommand{\distpct}{d}

\newcommand{\distpcto}{\distpct_{\perp}}

\newcommand{\bps}{\varsigma}

\newcommand{\bpe}{\epsilon}

\newcommand{\sigmaerror}{d_{\rm err}}

\newcommand{\sigmaerrorinfty}{d_{\infty}}

\newcommand{\uvpc}{\mathbf{\hat{r}}_{ij}}

\newcommand{\unitr}{\mathbf{\hat{e}}_{r}}

\newcommand{\vpc}{\mathbf{r}_{ij}}

\newcommand{\unitphi}{\mathbf{\hat{e}}_{\Aap}}

\newcommand{\unitAap}{\mathbf{\hat{e}}_{\Aap}}

\newcommand{\dpc}{r_{ij}}

\newcommand{\po}{\mathbf{\hat{o}}}

\newcommand{\prf}{\mathbf{\hat{o}}}

\newcommand{\cs}{A_c}

\newcommand{\dcs}{d\cs}

\newcommand{\pvf}{C}

\newcommand{\pcf}{\chi}

\newcommand{\pcfemuav}{\check{\chi}_{\rm av}}

\newcommand{\pcfref}{\chi_0}

\newcommand{\imsemu}{\check{I}}

\newcommand{\pn}{N_{\rm p}}

\newcommand{\pms}{M_{\rm par}}

\newcommand{\pv}{\vec{u}_{\rm par}}

\newcommand{\pacc}{\dot{\vec{u}}_{\rm par}}

\newcommand{\Fp}{\vec{F}}

\newcommand{\factorC}{{\cal C}}

\newcommand{\Ffe}{F_{\rm cap}}

\newcommand{\Fcap}{\Ffe}

\newcommand{\Fcappoint}{F_{\rm point}}

\newcommand{\Fcappointpm}{F_{\rm cap,point,\pm}}
\newcommand{\Fcappointm}{F_{\rm cap,point,-}}
\newcommand{\Fcappointp}{F_{\rm cap,point,+}}
\newcommand{\dFcappoint}{F_{\Delta,\rm cap,point}}

\newcommand{\Fdrag}{F_{\rm drag}}

\newcommand{\Fdragrot}{F_{\rm drag,rot}}

\newcommand{\efffreq}{\alpha_{\rm HO}}

\newcommand{\Fdragi}{F_{D}}

\newcommand{\Fdragiip}{F_{D,I}}

\newcommand{\Fdragib}{F_{D,b}}

\newcommand{\Cdrag}{c_{D}}

\newcommand{\Cdragm}{c_{DM}}

\newcommand{\Cdragbulk}{c_{\rm D,b}}

\newcommand{\Cdragif}{c_{\rm D,i}}

\newcommand{\DampCdrag}{\Cdragm}

\newcommand{\federkonst}{k_{\rm cap}}

\newcommand{\federkonstallg}{\federkonst}

\newcommand{\Freq}{\omega}

\newcommand{\lambdaho}{\lambda}
\newcommand{\omegaO}{\omega_0}
\newcommand{\omegaHO}{\omega_{a}}

\newcommand{\HOfita}{a_{\rm HO}}
\newcommand{\HOfitb}{b_{\rm HO}}
\newcommand{\HOfitc}{c_{\rm HO}}
\newcommand{\HOfitexpb}{d_{\rm HO}}

\newcommand{\mccp}{C_0}

\newcommand{\aIFDEFres}{a_{\rm res}}

\newcommand{\tpifdef}{t_p}
\newcommand{\tpifdefres}{{\tpifdef}_{\rm res}}

\newcommand{\ttranseff}{t_{\rm sp}}

\newcommand{\timeeq}{t_e}

\newcommand{\adsogeom}{\varepsilon}

\newcommand{\adsogeomconst}{{\rm const}}

\newcommand{\adsogeomM}{\varepsilon_M}

\newcommand{\cFEct}{c_{\rm dt}}
\newcommand{\ctFEct}{\tilde{c}_{\rm dt}}

\newcommand{\Crotdrag}{c_{\rm D,rot}}

\newcommand{\torqueconstjanus}{c_{\rm cap,dt,j}}

\newcommand{\Crotdragratio}{\gamma_{rot}}

\newcommand{\freqnullrot}{\omega_{\rm 0,rot}}

\newcommand{\mip}{I}

\newcommand{\Tp}{\vec{D}}

\newcommand{\Tpie}{\Tp_i}

\newcommand{\Tpieop}{\Tp_{i,1p}}

\newcommand{\Tpieinter}{\Tp_{\rm i, inter}}

\newcommand{\prot}{\vec{\omega}_{\rm par}}

\newcommand{\protacc}{\dot{\vec{\omega}}_{\rm par}}

\newcommand{\FSCcc}{\vec{F}^c}

\newcommand{\opS}{S}

\newcommand{\opSp}{\opS_{\parallel}}

\newcommand{\opQ}{Q}

  \newcommand{\Pcf}{g}

\newcommand{\acfl}{g_{2l}}

\newcommand{\acf}{g_{2}}

\newcommand{\pcfnorm}{\Pcf_n}

\newcommand{\ifa}{A_I}

\newcommand{\ifl}{L_I}

\newcommand{\sv}{V_S}

\newcommand{\sL}{L_S}

\newcommand{\pifa}{A_P}

\newcommand{\pa}{\vartheta}

\newcommand{\paO}{\pa_0}

\newcommand{\paOO}{\paOP}

\newcommand{\paOI}{\paOU}

\newcommand{\paOP}{\pa_{\rm PIO}}

\newcommand{\paOU}{\pa_{\rm UIO}}

\newcommand{\pamaxIst}{\pa_{\rm 1st,max}}

\newcommand{\dpaIst}{\Delta\pa_{\rm 1st,max}}

\newcommand{\pafin}{\pa_{\rm final}}

\newcommand{\pafinjpup}{\pa_{\rm up,Janus}}

\newcommand{\pafinuwup}{\pa_{\rm up,uw}}

\newcommand{\pafinup}{\pa_{\rm up}}
\newcommand{\pafinupr}{\pa_{\rm upright}}
\newcommand{\pafinmiddle}{\pa_{\rm middle}}

\newcommand{\pafinupoccif}{\pa_{\rm upU}}

\newcommand{\padamptransori}{\pa_{\rm dt,ori}}

\newcommand{\dotpa}{a}

\newcommand{\torquepa}{b}

\newcommand{\paor}{\pa_{or}}

\newcommand{\pam}{\pa_{m}}

\newcommand{\paeff}{\pa'}

\newcommand{\paeffO}{\pa_0'}

\newcommand{\paeffOn}{n_\pa}

\newcommand{\paefffin}{\pa_{\rm final}'}

\newcommand{\paefffinjpup}{\pa_{\rm Jup}'}

\newcommand{\paefffinuwup}{\pa_{\rm up,uw}'}

\newcommand{\paefffinup}{\pa_{\rm up}'}

\newcommand{\patmp}{\pa_{\rm tmp}}

\newcommand{\paens}{\pa_{\rm ens}}

\newcommand{\paense}{\pa_e}

\newcommand{\Aa}{\phi}

\newcommand{\Aap}{\phi_p}

\newcommand{\mpo}{\tilde{m}}

\newcommand{\Aapr}{\Delta\phi_{\mpo}}

\newcommand{\iu}{h}

\newcommand{\iuA}{A}

\newcommand{\iuAO}{\iuA_0}

\newcommand{\iuAOfit}{a_0}

\newcommand{\iuAOmfit}{a_m}

\newcommand{\iuAtilde}{\tilde{A}}

\newcommand{\iuB}{B}

\newcommand{\spv}{\radint}

\newcommand{\spvO}{\spv_0}

\newcommand{\spvsep}{\spv}

\newcommand{\Par}{m}

\newcommand{\Parzd}{\Par_{2d}}

\newcommand{\Ecc}{{\cal E}}

\newcommand{\Gaf}{G}

\newcommand{\Vp}{V_p}

\newcommand{\FE}{{\cal F}}

\newcommand{\FEeq}{\FE_{\rm eq}}

\newcommand{\TEP}{e}

\newcommand{\FEI}{\FE_0}

\newcommand{\FEPA}{\FE_{\parallel}}

\newcommand{\FEPE}{\FE_{\perp}}

\newcommand{\FEPAE}{\FE_{\parallel,eq}}

\newcommand{\FEPEE}{\FE_{\perp,eq}}

\newcommand{\FEPAJ}{\FE_{\parallel}}

\newcommand{\FEPEJ}{\FE_{\perp}}

\newcommand{\FEEMU}{\check{\FE}}

\newcommand{\FEIFEMU}{\check{\FE}_{\rm if}}

\newcommand{\FEPEMU}{\check{\FE}_{\rm pressure}}

\newcommand{\FEINTEREMU}{\check{\FE}_{\rm interaction}}

\newcommand{\FEENTREMU}{\check{\FE}_{\rm entr}}

\newcommand{\ENTREMU}{\check{\entr}}

\newcommand{\FEEMUS}{\FE_S}

\newcommand{\FEENTREMUS}{\FE_{\rm entr}}

\newcommand{\FEIFEMUS}{\FE_{\rm if}}

\newcommand{\FEG}{{\cal G}}

\newcommand{\dFEG}{\Delta\Gamma}

\newcommand{\sts}{\ST_{\rm red}}

\newcommand{\stUD}{\ST_{12}}

\newcommand{\iaemus}{\ia_{12}}

\newcommand{\volfracdp}{\Phi}

\newcommand{\ENTREMUS}{\entr_S}

\newcommand{\ndropemus}{n_{D}}

\newcommand{\fesaf}{{\cal A}_\pid}

\newcommand{\FENUC}{\FE_{\rm nuc}}

\newcommand{\FEMC}{\FE_{\rm MC}}

\newcommand{\FEMCB}{\psi_{\rm MC}}

\newcommand{\FEMCIF}{\kappa_{\rm MC}}

\newcommand{\FEdt}{\FE_{dt}}

\newcommand{\dFEdt}{\Delta\FE_{dt}}

\newcommand{\dFEdtpm}{\Delta\FE_{dt\pm}}

\newcommand{\dFEdtp}{\Delta\FE_{dt+}}

\newcommand{\dFEdtm}{\Delta\FE_{dt-}}

\newcommand{\dFEdtd}{\Delta\FE_{dt\Delta}}

\newcommand{\ST}{\sigma}

\newcommand{\gbr}{g_{br}}

\newcommand{\lt}{\kappa}

\newcommand{\rlt}{\tilde{\lt}}

\newcommand{\ia}{A}

\newcommand{\iaO}{\ia_{\rm flat}}

\newcommand{\iap}{\ia_{\parallel}}

\newcommand{\iao}{\ia_{\perp}}

\newcommand{\iaOemu}{\check{\ia}_{\rm 12,0}}

\newcommand{\iaemu}{\check{\ia}}

\newcommand{\iaex}{\ia_{\rm ex}}

\newcommand{\iaexflat}{\ia_{\rm ef}}

\newcommand{\iaexemuav}{\check{\ia}_{\rm ex}}

\newcommand{\iapemu}{\check{\ia}_{\rm 12,p}}

\newcommand{\iaidealundef}{\ia_{\rm undef}}

\newcommand{\iaidealundefex}{\ia_{\rm ex,undef}}

\newcommand{\iaSell}{\ia_{\rm ell}}

\newcommand{\iaSellh}{\ia_{ellh}}

\newcommand{\iafell}{a_{\rm ell}}

\newcommand{\iaR}{\overline{\ia}}

\newcommand{\iapr}{\ia_{\rm pr}}

\newcommand{\iaunpr}{\ia_{\rm unpr}}

\newcommand{\iadeltaif}{\ia_{\rm ifdef}}
\newcommand{\Diadeltaif}{\Delta\iadeltaif}

\newcommand{\ratioifdefundef}{\Xi}

\newcommand{\DiadeltaifAI}{A_1}
\newcommand{\DiadeltaifBO}{B_0}
\newcommand{\DiadeltaifBI}{B_1}
\newcommand{\DiadeltaifBII}{B_2}

\newcommand{\iatilde}{\tilde{\ia}}

\newcommand{\iatildeexflat}{\iatilde_{\rm ef}}

\newcommand{\iatildepr}{\iatilde_{\rm pr}}

\newcommand{\iatildeunpr}{\iatilde_{\rm unpr}}

\newcommand{\iatildeSell}{\iatilde_{\rm ell}}

\newcommand{\tpcl}{{\cal L}}

\newcommand{\tpclemu}{\check{\tpcl}}

\newcommand{\dip}{\xi}

\newcommand{\dipO}{\dip_0}

\newcommand{\dipOa}{\dip_{a}}

\newcommand{\dipfinal}{\dip_{\rm final}}

\newcommand{\dipfinalup}{\dip_{\rm up}}

\newcommand{\dipfinunpr}{\dip_{\rm unpr}}

\newcommand{\dipfinalpr}{\dip_{\rm pr}}

\newcommand{\dips}{\dip_s}

\newcommand{\tdips}{t_{\dip,s}}

\newcommand{\dipJpairsplit}{\dip_{\rm split}}

\newcommand{\dipJSLiftouch}{\dip_{\rm JSL,if}}

\newcommand{\dipJSLiftouchpcoord}{\dip_{\rm JSL,if,pc}}

\newcommand{\dipJSLiftouchifdef}{\dip_{\rm JSL,if,def}}

\newcommand{\piid}{\tilde{\xi}}

\newcommand{\diptreff}{\dip_{\rm etrans}}

\newcommand{\diptreffuw}{\dip_{\rm trans,eff}}

\newcommand{\diptmp}{\dip_{\rm tmp}}

\newcommand{\dpi}{z}

\newcommand{\pid}{z}

\newcommand{\idef}{\lambda}

\newcommand{\idefti}{\Lambda}

\newcommand{\ii}{y_{\rm if,0}}

\newcommand{\ri}{y_{\rm if}}

\newcommand{\ift}{d}

\newcommand{\startingpointA}{{|\dip(t=0)|}}

\newcommand{\dxpi}{x}

\newcommand{\dypi}{z}

\newcommand{\elSA}{S}

\newcommand{\IFDEF}{\iu}

\newcommand{\gsqifdef}{\Upsilon}
\newcommand{\gsqifdefvor}{\Upsilon_C}
\newcommand{\gsqifdefr}{\Upsilon_\spv}
\newcommand{\gsqifdefphip}{\Upsilon_{\Aap}}
\newcommand{\gsqifdefphipt}{\tilde{\Upsilon}_{\Aap}}
\newcommand{\gsqifdefm}{\gsqifdef_{\Par}}
\newcommand{\gsqifdefmJ}{\gsqifdef_{\Par,\gsqifdefmJJ}}
\newcommand{\gsqifdefmJJ}{j}
\newcommand{\gsqifdefmO}{\gsqifdef_{\Par,0}}
\newcommand{\gsqifdefmI}{\gsqifdef_{\Par,1}}
\newcommand{\gsqifdefmII}{\gsqifdef_{\Par,2}}
\newcommand{\gsqifdefmIII}{\gsqifdef_{\Par,3}}
\newcommand{\gsqifdefmIV}{\gsqifdef_{\Par,4}}
\newcommand{\gsqifdefmV}{\gsqifdef_{\Par,5}}
\newcommand{\gsqifdefmVI}{\gsqifdef_{\Par,6}}
\newcommand{\gsqifdefmVII}{\gsqifdef_{\Par,7}}
\newcommand{\gsqifdefmVIII}{\gsqifdef_{\Par,8}}
\newcommand{\gsqifdefmIX}{\gsqifdef_{\Par,9}}
\newcommand{\gsqifdefell}{\Upsilon_{\rm ell}}
\newcommand{\gsqifdefnum}{\Upsilon_{\rm num}}
\newcommand{\gsqifdefintphip}{\Upsilon_{{\rm int}}}

\newcommand{\ifdefatp}{\Delta\dip}

\newcommand{\ifdefatpfitdipG}{H}

\newcommand{\ifdefatpfitdipO}{\ifdefatpfitdipG_0}

\newcommand{\ifdefatpfitdipI}{\ifdefatpfitdipG_1}

\newcommand{\ifdefatpfitdipII}{\ifdefatpfitdipG_2}

\newcommand{\lpdxpi}{i_x}

\newcommand{\lpdypi}{i_z}

\newcommand{\lpdg}{i_j}

\newcommand{\nlp}{N}

\newcommand{\nlpx}{\nlp_x}

\newcommand{\nlpy}{\nlp_y}

\newcommand{\nlpz}{\nlp_z}

\newcommand{\nlpxpi}{\nlp_x}

\newcommand{\nlpypi}{\nlp_z}

\newcommand{\nlpg}{\nlp_j}

\newcommand{\Doi}{y}

\newcommand{\nfidef}{\lambda_p}

\newcommand{\tlb}{t}

\newcommand{\tres}{t}

\newcommand{\Fsc}{\vec{F}^c}

\newcommand{\mwf}{\Psi^c}

\newcommand{\mwfII}{\Psi^{c'}}

\newcommand{\ccfsg}{g}

\newcommand{\ccfsgccs}{\ccfsg_{cc'}}

\newcommand{\ccfsgcc}{\ccfsg_{cc}}

\newcommand{\ccfsgbr}{\ccfsg_{br}}

\newcommand{\ds}{L(t)}

\newcommand{\DS}{L}

\newcommand{\dsav}{L}

\newcommand{\DSi}{\DS_i}

\newcommand{\stf}{s}

\newcommand{\stfn}{\stf_n}

\newcommand{\wsf}{k}

\newcommand{\wvsf}{\vec{k}}

\newcommand{\sfpr}{\Upsilon}

\newcommand{\spdf}{f}

\newcommand{\spdfeq}{\spdf^{\mathrm{eq}}}

\newcommand{\spdfi}{f_i}

\newcommand{\spdfieq}{\spdfi^{\mathrm{eq}}}

\newcommand{\colfreq}{\omega}

\newcommand{\relaxt}{\tau}

\newcommand{\dtimeLB}{\Delta t}

\newcommand{\dxLB}{\Delta x}

\newcommand{\colBGK}{\Omega}

\newcommand{\colBGKi}{\colBGK_i}

\newcommand{\mrtrtm}{{\cal S}}

\newcommand{\mrtmm}{{\cal M}}

\newcommand{\lpx}{\mathbf{x}}

\newcommand{\nflnNFN}{N_{\mathrm{FN}}}

\newcommand{\lpxalt}{\mathbf{x}}

\newcommand{\gpp}{\vec{x}}

\newcommand{\gppalt}{\vec{x}}

\newcommand{\gpv}{\vec{v}}

\newcommand{\gpvi}{v_i}

\newcommand{\fgv}{\vec{u}}

\newcommand{\fgvi}{u_i}

\newcommand{\tgpv}{\vec{c}}

\newcommand{\gpf}{\vec{F}_p}

\newcommand{\gpm}{M_{p}}

\newcommand{\tzd}{n}

\newcommand{\lvci}{\vec{c}_i}

\newcommand{\LBc}{c}

\newcommand{\dens}{\rho}

\newcommand{\densc}{\rho^{\LBc}}

\newcommand{\denst}{\rho^{\rm total}}

\newcommand{\densb}{\rho^b}

\newcommand{\densr}{\rho^r}

\newcommand{\densO}{\rho_0}

\newcommand{\dynaVisc}{\mu}

\newcommand{\kineVisc}{\nu}

\newcommand{\CaNu}{Ca}

\newcommand{\Kn}{Kn}

\newcommand{\CaNuMa}{\CaNu_{\rm max}}

\newcommand{\CaNuAv}{\CaNu_{\rm av}}

\newcommand{\densnuc}{\rho_{\rm nuc}}

\newcommand{\cpnuc}{\Delta\mu_{\rm nuc}}

\newcommand{\spos}{c_s}

\newcommand{\Ma}{Ma}

\newcommand{\Mav}{v_s}

\newcommand{\fop}{\varphi}

\newcommand{\fopft}{\tilde{\fop}}

\newcommand{\fopftfl}{\fop'}

\newcommand{\ffr}{ffr}

\newcommand{\lcf}{\mathbf{F}_{ij}}

\newcommand{\lcfda}{\mathbf{\tilde{F}}_{ij}}

\title{Equilibrium orientation and adsorption of an ellipsoidal Janus particle at a fluid-fluid interface}
\author{Florian G\"unther}
\email{fguenther@eclipso.de}
\affiliation{Department of Applied Physics, Eindhoven University of Technology, P.O. Box 513, 5600MB Eindhoven, The Netherlands}
\author{Qingguang Xie}
\email{q.xie1@tue.nl}
\affiliation{Department of Applied Physics, Eindhoven University of Technology, P.O. Box 513, 5600MB Eindhoven, The Netherlands}
\author{Jens Harting}
\email{j.harting@fz-juelich.de}
\affiliation{Helmholtz Institute Erlangen-N\"urnberg for Renewable Energy (IEK-11), Forschungszentrum J\"ulich, F\"urther Str. 248, 90429 N\"urnberg, Germany}
\affiliation{Department of Chemical and Biological Engineering and Department of Physics, Friedrich-Alexander-Universit\"at Erlangen-N\"urnberg, F\"urther Str. 248, 90429 N\"urnberg, Germany}
\date{\today}
\begin{abstract}
We investigate the equilibrium orientation and adsorption process of a single,
ellipsoidal Janus particle at a fluid-fluid interface. The particle surface comprises
equally sized parts that are hydrophobic or hydrophilic. We present free energy
models to predict the equilibrium orientation and compare the theoretical
predictions with lattice Boltzmann simulations. We find that the deformation of
the fluid interface strongly influences the equilibrium orientation of the
Janus ellipsoid.  The adsorption process of the Janus ellipsoid can lead to
different final orientations determined by the interplay of particle aspect
ratio and particle wettablity contrast.

\end{abstract}
\maketitle

\section{Introduction}
\label{sec:JanusIntroAllgUnd1p}
Janus particles have drawn great attention recently for their potential in
various applications, such as micro-swimmers, stabilizers of emulsions and
catalysis~\cite{Walther2013a,FernandezRodriguez2014a,FernandezRodriguez2016a,Archer2018,Marschelke2020}.
The special characteristics of these particles include their anisotropic
physical (e.g. optical, electric, or magnetic) or chemical (e.g. wetting or
catalytic) properties at well-defined areas on their surface.
Amphiphilic Janus particles are characterized by opposite wetting abilities at
their hemispheres and have been shown to attach strongly at fluid-fluid
interfaces~\cite{Morris2015a,Park2012a,Park2012b,Hirose2007a,Binks2001b}. Binks
et al.~\cite{Binks2001b} compared the free energy of homogeneous spherical
particles and Janus spherical particles in contact with a fluid interface. They
found that the free energy of amphiphilic Janus particles can be $3$ times
larger than that of homogeneous particles. Furthermore, Janus particles retain
strong adsorption at interfaces even for contact angles of $0^{\circ}$ and
$180^{\circ}$.  Thus, Janus particles are generally expected to be more
efficient emulsion stabilizers than homogeneous colloidal
particles~\cite{Fan2012,Mejia2012,Hirose2007a}.

The understanding of the equilibrium orientation and the adsorption dynamics of
Janus particles at fluid interfaces is critical to enable their optimal
utilization.  Park et al. theoretically studied the configurations of Janus
ellipsoids at a fluid interface~\cite{Park2012a, Park2012b}, and found that the
orientation of ellipsoidal Janus particles is affected by the particle aspect
ratio and the Janus character given by the wettability contrast. In their
model, they used a flat-interface approach, where the dynamic deformation of
interface during the adsorption process is neglected. However, numerical
studies of ellipsoidal Janus particles at fluid interfaces demonstrated that
the interface deforms around the particle and thus, the interface deformation
is expected to affect the equilibrium orientation of
particles~\cite{Rezvantalab2013a, Hirose2007a, Anzivino2019, XieHarting2020}.

Here, we present a simplified free energy model of an ellipsoidal Janus particle at a
non-deforming fluid-fluid interface and then extend it by taking into account the interface deformation. The
theoretical models are used to predict the equilibrium orientation of the
particle, and are compared to our numerical results obtained from lattice
Boltzmann simulations. Moreover, we numerically study the adsorption process
of a Janus ellipsoid with varying the particle aspect ratio and the Janus character,
i.e. the wettability contrast.

The remainder of this paper is organized as follows. In
Sec.~\ref{sec:sim-method} we introduce our simulation method and setup,
followed by a comparison of the simplified free energy model with simulations
in Sec.~\ref{sec:SJMFA}.  Sec.~\ref{sec:FE1JpFull} extends towards
deformable interfaces. In Sec.~\ref{sec:adsorption} we demonstrate how the
adsorption trajectories of an ellipsoidal Janus particle depend on the
interplay between particle aspect ratio and wettability contrast. Finally, we
conclude the paper with a short summary.

\section{Simulation method}
\label{sec:sim-method}
Here, we summarize the main ingredients of our simulation method. A more
detailed description can be found in our previous publications on
particle-laden multicomponent
flows~\cite{Frijters2012a,Guenther2012a,Krueger2012c,Guenther2013a,Guenther2013b,Guenther2014a}.
\label{ssec:lb}
We use the lattice Boltzmann (LB) method~\cite{succi2001a} to simulate two
immiscible fluids $c$, $c'$ where the discrete form of the Boltzmann equation
can be written as
\begin{equation}
  \label{eq:LBG}
  \spdfi^c(\lpx + \lvci \Delta t , t + \Delta t)=\spdfi^c(\lpx,t)+\Omega_i^c(\lpx,t)
  \mbox{.}
\end{equation}
Here, $\spdfi^c(\lpx,t)$ is the single-particle distribution function for fluid
component $c$ with discrete lattice velocity $\lvci$ at time $t$ located
at lattice position $\lpx$. We use a D3Q19 lattice, i.e. a 3D lattice with a lattice constant
$\Delta x$ and nineteen velocity
directions. $\Delta t$ is the
timestep and 
\begin{equation}
  \label{eq:BGK_collision_operator}
  \Omega_i^c(\lpx,t) = -\frac{\spdfi^c(\lpx,t)- \spdfi^\mathrm{eq}(\rho^c(\lpx,t), \vec{u}^c(\lpx,t))}{\left( \tau^c / \Delta t \right)}
\end{equation}
is the Bhatnagar-Gross-Krook (BGK) collision
operator~\cite{bhatnagar1954a}. The density is defined as
\mbox{$\rho^c(\lpx,t)=\rho_0\sum_i\spdfi^c(\lpx,t)$}, where $\rho_0$ is a reference density.
$\tau^c$ is the relaxation time for the component $c$ and
$\spdfi^{\mathrm{eq}}(\rho^c,\vec{u}^c)$ is the equilibrium distribution function. 
$\vec{u}^c(\lpx,t)=\sum_i\spdfi^c(\lpx,t)\lvci/\rho^c(\lpx,t)$ is the local velocity.
The kinematic viscosity can be calculated as
\begin{equation}
  \label{eq:kinvis}
  \nu^c = c_s^2 \Delta t \left( \frac{\tau^c}{\Delta t} - \frac{1}{2} \right)
  \mbox{,}
\end{equation}
where $c_s=\frac{1}{\sqrt{3}}\frac{\Delta x}{\Delta t}$
is the speed of sound.
In the following, we choose $\Delta x = \Delta t = \rho_0 = 1$ for
simplicity.  Furthermore, in all simulations, the relaxation time is set to $\tau^c \equiv 1$.

\label{ssec:multicomponent-lb}
We use the method introduced by Shan and Chen for the simulation of a multicomponent fluid mixture~\cite{shan1993a}. Every species
has its own distribution function following \eqnref{LBG} and an interaction between the different components is introduced as 
\begin{equation}
  \label{eq:sc}
  \Fsc(\lpx,t) = -\mwf(\lpx,t) \sum_{c'}g_{cc'} \sum_{\lpx'} \Psi^{c'}(\lpx',t) (\lpx'-\lpx)
  \mbox{.}
\end{equation} 
The resulting force is included in the equilibrium distribution function
(\eqnref{BGK_collision_operator}) by shifting the velocity $\vec{u}^c(\lpx,t)$.
$g_{cc'}$ is the interaction parameter between the fluid components $c$ and
$c'$ and defines the surface tension and $\Psi^c(\mathbf{x}, t) = \rho_0 [1-\exp({-\rho^c(\mathbf{x},t)/\rho_0})]$
is a functional defining the equation of state of the system. 

\label{ssec:nanoparticles}
The particles follow Newton's equations of motion and are discretized on the
lattice. They are coupled to both fluid species by a modified bounce-back
boundary condition which was originally introduced by
Ladd~\cite{Jansen2011a,aidun1998a,ladd1994a,ladd1994b,ladd2001a,KHH04}.
If the particle moves, some lattice nodes become free and others become
occupied. The momentum of the fluid on the newly occupied nodes is transferred
to the particle.
A newly freed node (located at $\lpx$) is filled with the average density of the $N_{\mathrm{FN}}$ neighboring fluid
lattice nodes $\lpx_{i_\mathrm{FN}}$ for each component $c$,
\begin{equation}
  \label{eq:rho-surr}
  \overline{\rho}^c(\lpx,t) \equiv \frac{1}{N_{\mathrm{FN}}} \sum_{i_\mathrm{FN}}  \rho^c(\lpx+\vec{c}_{i_{\mathrm{FN}}},t)
  \mbox{.}
\end{equation}
The fluid interaction forces also act between a node in the outer shell of a particle
and its neighboring point outside of the particle. Since this would lead to an increase
of the fluid density around the particle, the nodes in the outer shell of the particle are filled with a virtual fluid
corresponding to the average of the value in the neighboring free nodes for
each fluid component:
$\rho_{\mathrm{virt}}^c(\lpx,t) = \overline{\rho}^c(\lpx,t)$.
This can be used to control the wettability properties of the particle
surface for the special case of two fluid species which will be named red and blue. We define the parameter $\Delta\rho$ and call it particle color. For
positive values of $\Delta\rho$ we add it to the fluid component $c$,
  $\rho_{\mathrm{virt}}^c=\overline{\rho}^c+\Delta\rho$, 
while for negative values we add its absolute value to another fluid component $c'$,
  $\rho_{\mathrm{virt}}^{c'}=\overline{\rho}^{c'}+|\Delta\rho|$.


\begin{figure}
	\centering
	\includegraphics[width=6cm]{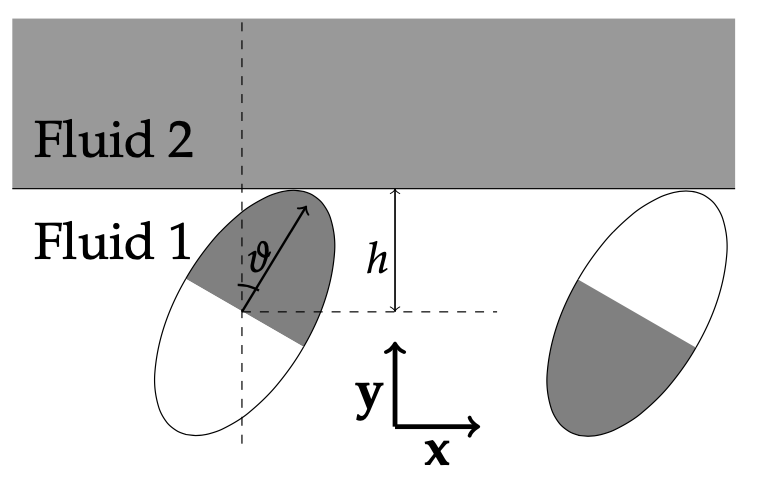}
	\caption{	A single ellipsoidal Janus particle touching a fluid-fluid interface:
		$h$ is the distance between the particle center and the flat fluid
		interface, $\pa$ is the particle orientation. 
		Fluid 1 is defined as the fluid in which the particle is immersed in the beginning
		and fluid 2 is the fluid into which the particle enters.
		The dark area of the Janus particle prefers fluid 2 whereas the white area
		prefers fluid 1. 
		The left particle is in the ``preferred initial orientation'' (PIO,
		$0\le\paO<90^{\circ}$) and the right particle in the ``unpreferred initial
		orientation'' (UIO, $90^{\circ}<\paO\le180^{\circ}$). 
	}
\label{fig:geo}
\end{figure}

Our simulation setup is illustrated in Fig.~\ref{fig:geo} and consists of a
cubic volume composed of two equally sized layers of two immiscible fluids such
as oil and water with density 0.7 each. With setting the coupling constant $g_{cc'}=0.1$, we obtain a
surface tension of $\ST_{12}\approx0.041$ and our fluids form a flat
fluid-fluid interface. The system is confined by walls at the top and the
bottom. Periodic boundary conditions apply in the $\dxpi$- and
$\dypi$-direction parallel to the interface.


We restrict ourselves to symmetric ellipsoidal Janus particles with an aspect
ratio $m$.  $\prp$ and $\pro$ are the parallel and orthogonal radius of the
ellipsoid, respectively (The parallel radius is chosen as $\prp=8$ if not
defined otherwise.).  The particle surface is divided into two equal
proportions: a more apolar region with a contact angle $\pcaa=90^{\circ}+\psjp$
and a more polar region with a contact angle $\pcap=90^{\circ}-\psjp$.  The
Janus influence is given by the parameter $\psjp$. Both areas of different
wettability have exactly the same size. The position and the orientation of the
particle with respect to the undeformed, flat interface are characterized by
the variables $\dip$ and $\pa$, respectively. $\dip$ is the distance between
the undeformed interface and the particle center in units of $\prp$,
$\dip=h/\prp$ and $\pa$ is the polar angle and given as the angle between the
main axis and the interface normal.  The particle is initially placed so that
it just touches the flat interface with different initial orientations
$\paO=\pa(t=0)$ and initial positions $\dipO=\dip(t=0)$. 

The particle can be placed in two different ways as shown in
Fig.~\ref{fig:geo}. It adsorbs to the interface between the fluids 1 and 2.
Here, fluid 1 is defined as the fluid where the particle is immersed at the
start of the simulation and fluid 2 is the fluid in which the particle enters
during the adsorption process. The Janus particle is divided into an area which
prefers fluid 1 (white area in Fig.~\ref{fig:geo}) and one preferring fluid 2
(dark area in Fig.~\ref{fig:geo}). This leads to two possible situations of the
initial particle orientation for the adsorbing process: the first orientation
range is the ``preferred initial orientation'' (PIO) with $0\le\paO<90^{\circ}$
(left particle in Fig.~\ref{fig:geo}), where the wetting area preferring fluid
2 touches the interface and enters fluid 2.  The next one is the ``unpreferred
initial orientation'' (UIO) with $90^{\circ}<\paO\le180^{\circ}$ (right
particle in Fig.~\ref{fig:geo}), where the oppositely wetting area is at the
interface.  In case of $\paO=90^{\circ}$, the border between both wetting areas
touches the interface.  $\paeff$ is defined as the effective polar angle and
given as
\begin{equation}
\label{eq:paeffdef}
\paeff=
\begin{cases}
\pa, & \text{for }\pa\le90^{\circ}\\
180^{\circ}-\pa, & \text{for }\pa>90^{\circ}
\mbox{.}
\end{cases}
\end{equation}
%
Five reference values for the effective initial orientation $\paeffO$ are chosen as
\begin{equation}
\label{eq:paeffO}
\paeffO=\paeffOn22.5^{\circ}\text{, }\paeffOn=\{0, 1, 2, 3, 4\}
\mbox{.}
\end{equation}
All these initial orientations $\paeffO$ are taken for the PIO and the UIO.
$\paeffO=(10^{-4})^{\circ}$ is chosen as an additional orientation and only useful
for the PIO as a hint if the upright orientation ($\pa=0$) is a minimum of the free energy. 

\section{A simplified free energy model for Janus particles}
\label{sec:SJMFA}\label{ssec:FullAnalytical}
In this section we present a simplified free energy model for a single Janus
particle at a flat fluid interface. The following two assumptions are made:
first, we consider an undeformable, flat interface. Second, only the extreme
particle orientations parallel and orthogonal to the interface are taken into
account.  The upright orientation is the one where each wetting area is fully
immersed in its preferred fluid. This orientation is expected if the Janus
effect dominates. The parallel orientation is the orientation, where the
particle occupies as much interfacial area as possible. This orientation is
expected if the shape of the particle dominates.  This model is referred to in
the following as the ``simplified Janus model''.  

The free energy of the system is given as
\begin{equation}
  \label{eq:FJol}
  \FE(\dip,\pa)=\ST_{12}\ia_{12}+\sum_{i=1,2}\sum_{j=A,P}\ST_{ij}\ia_{ij}
  \mbox{.}
\end{equation}
$\ia$ and $\ST$ are the interface areas and the interfacial tensions,
respectively. 1 and 2 denote the two fluids whereas $P$ and $A$ denote the
polar and apolar region of the particle surface, respectively.

\begin{figure}
\centering
\includegraphics[width=7cm]{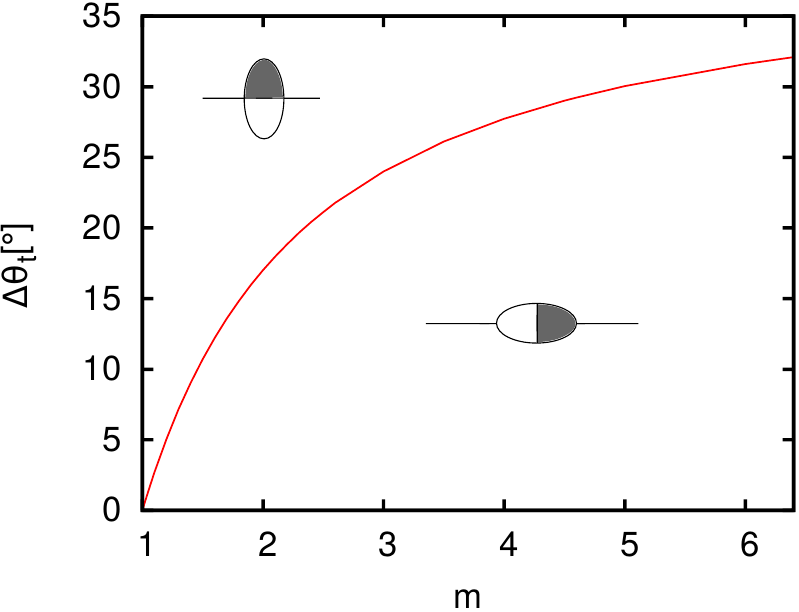}
\caption{The phase diagram ($\psjpt$-$\Par$) obtained from the ``simplified
	Janus model''. In the area above the line the upright orientation is
	the stable state whereas ibelow the line the particle orientation
	parallel to the interface is the stable state. The line represents
	\eqnref{JRelm}.
}
\label{theo1predSJ}
\end{figure}

We consider the initial state where the particle is fully immersed in fluid 1
which changes \eqnref{FJol} to
\begin{equation}
  \label{eq:FEJI}
	\FEI=\ST_{12}\iaO+\frac{1}{2}\iaSell(\ST_{1A}+\ST_{1P})
  \mbox{.}
\end{equation}
$\iaSell$ is the total area of the ellipsoid surface.  $\iaO$ is the area of
the flat fluid interface in absence of the particle.  $\FEI$ is the free energy
of the reference state  where the particle is immersed in the bulk and away
from the interface.  For the final particle orientation parallel to the
interface
  ($\pafin=90^\circ$), \eqnref{FJol} changes to
\begin{equation}
  \label{eq:FPx}
\FEPAJ=(\iaO-\iap)\ST_{12}+0.25\iaSell\left(\sum_{i=1,2}(\ST_{iP}+\ST_{iA})\right)
  \mbox{.}
\end{equation}
$\iaO-\iap$ is the remaining fluid-fluid interface after the particle
adsorption and $\iap$ is the area of the fluid-fluid interface excluded by the
particle oriented parallel to the interface.  The free energy difference is
given as
$\Delta\FEPAJ=\FEPAJ-\FEI=-\iap\ST_{12}+0.25\iaSell(\sum_{i=A,P}(\ST_{2i}-\ST_{1i}))$. 
Using Young's equation $\sigma_{12}\cos\theta=\sigma_{2p}-\sigma_{1p}$
which holds for both of the two wetting areas of the particle independently, we
obtain
\begin{equation}
  \label{eq:FadPx}
\Delta\FEPAJ=\ST_{12}\left(0.25\iaSell\left(\sum_{i=A,P}\cos\pca_i\right)-\iap\right)
  \mbox{.}
\end{equation}
Furthermore, assuming $\cos\pcaa=-\cos\pcap$ to simplify \eqnref{FadPx}, we get
\begin{equation}
  \label{eq:FadP}
\Delta\FEPAJ=-\ST_{12}\iap
  \mbox{.}
\end{equation}
For the particle orientation parallel to the interface normal, the free energy
is given as
$\FEPEJ=(\iaO-\iao)\ST_{12}+0.5\iaSell(\ST_{1A}+\ST_{2P})$.
$\iao$ is the excluded interfacial area due to a particle with orthogonal orientation.
In the same way as above, we get
$\Delta\FEPEJ=\FEPEJ-\FEI=-\iao\ST_{12}+0.5\iaSell(\ST_{2P}-\ST_{1P})$ and reach the
following relation:
\begin{equation}
  \label{eq:FadN}
\Delta\FEPEJ=-\ST_{12}(0.5\iaSell\cos\pcap+\iao) =-\ST_{12}(0.5\iaSell\sin\psjp+\iao)
  \mbox{,}
\end{equation}
with $\cos\pcap=\sin\psjp$.
In order to find the orientation of the stable point of the Janus particle at the
  interface, it is necessary to find the global minimum of the free energy.
$\psjpt$ is defined as the transition value of $\psjp$ above
which the particle orientation parallel to the interface normal minimizes the
free energy.
Comparing \eqnref{FadP} and \eqnref{FadN} leads to the following value of $\psjpt$ for the transition point:
\begin{equation}
  \label{eq:JRelA}
\sin\psjpt=\frac{2(\iap-\iao)}{\iaSell}
  \mbox{.}
\end{equation}
Thus, $\sin\psjpt$ depends on the aspect ratio via the three interfacial
areas, which are given as
\begin{equation}
\begin{split}
  \label{eq:sfajp}
  0.5\iaSell&=\pi \pro^2+\frac{\pi
  \pro\prp^2}{\sqrt{\prp^2-\pro^2}}\arcsin\left(\frac{\sqrt{\prp^2-\pro^2}}{\prp}\right),\\
      &=\pi \pro^2\left(1+\frac{\Par}{\sqrt{\Par^2-1}}\arcsin\left(\frac{\sqrt{\Par^2-1}}{\Par}\right)\right)\mbox{,}\\
      \iao&=\pi\pro^2, \\ 
       \iap&=\pi \pro\prp=\pi\Par \pro^2
  \mbox{.}
\end{split}
\end{equation}
This changes \eqnref{JRelA} to the following relation for $\psjpt(\Par)$:
\begin{equation}
  \label{eq:JRelm}
\sin\psjpt=\frac{\Par-1}{\left(1+\frac{\Par}{\sqrt{\Par^2-1}}\arcsin\left(\frac{\sqrt{\Par^2-1}}{\Par}\right)\right)}
  \mbox{.}
\end{equation}
This result relates the Janus parameter at the transition point to the aspect
ratio of the particle. As an example, for $\Par=2$, we obtain
$\psjpt\approx17^{\circ}$.  The function given in \eqnref{JRelm} is shown in
Fig.~\ref{theo1predSJ} for the range of the aspect ratio which is used in the
simulation of the particle adsorption below ($1\le\Par\le6$). The parallel
orientation minimizes the free energy below the line and the upright
orientation above.

\begin{figure}
\centering
\includegraphics[width=7cm]{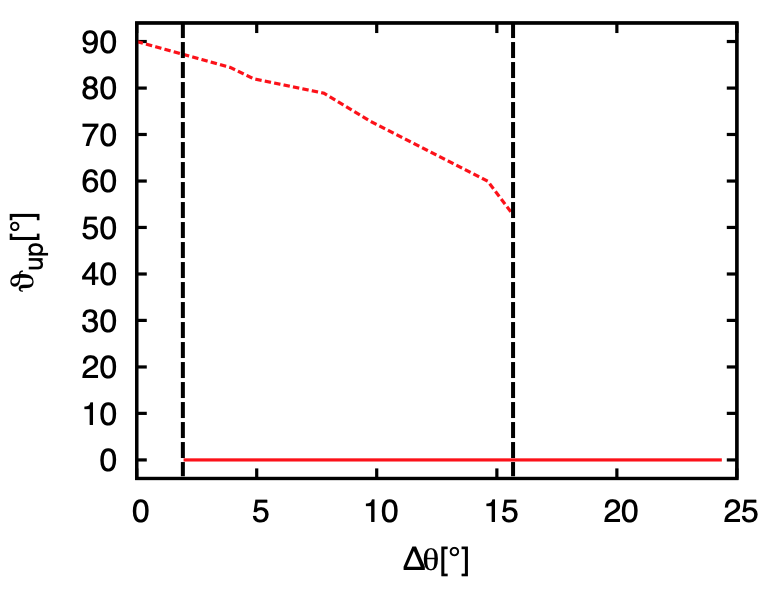}
\caption{Final orientation of a Janus particle for $\Par=1.5$.  A tilted
	(red dashed line) and an upright (red solid line) orientation of the particle
	are shown depending on the Janus parameter $\psjp$.  The vertical dashed lines
	divide the range of $\psjp$ into three regions. In the left region, it
	can be concluded from the simulation results that the tilted
	orientation is the global minimum.  In the right region, the upright
	orientation is the only one which is found.  In the central region,
	both tilted and upright orientations are found in the simulation
	results. It cannot be seen which one corresponds to the energetic
	minimum.  
}
\label{dendpunktcolvglmuprOri}
\end{figure}
In the following, we compare the predictions from the simplified model to our
simulations. We follow several paths through the $\Par$-$\psjp$ phase diagram
from Fig.~\ref{theo1predSJ} and compare the equilibrium orientation with the
final particle orientations obtained from the simulation.  At first, we follow
the lines along a constant aspect ratio $\Par$ with a varying Janus parameter
$\psjp$.  This corresponds to a vertical line through the phase diagram.
\begin{figure}
  \centering
  \subfigure[]
  {\label{dendpunktcolvglm}\includegraphics[width=7cm]{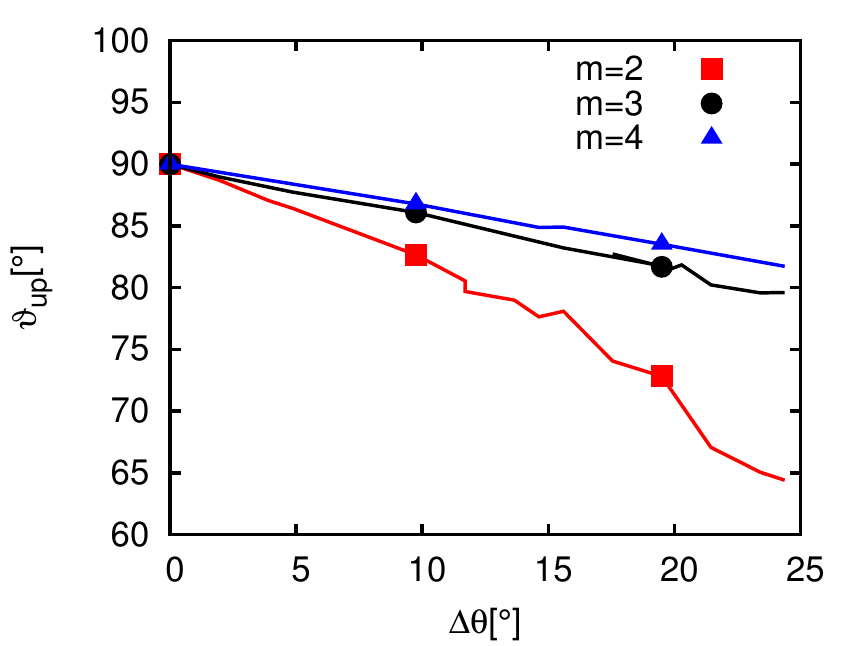}}
\qquad
  \subfigure[]
  {\label{dendpunktmdep}\includegraphics[width=7cm]{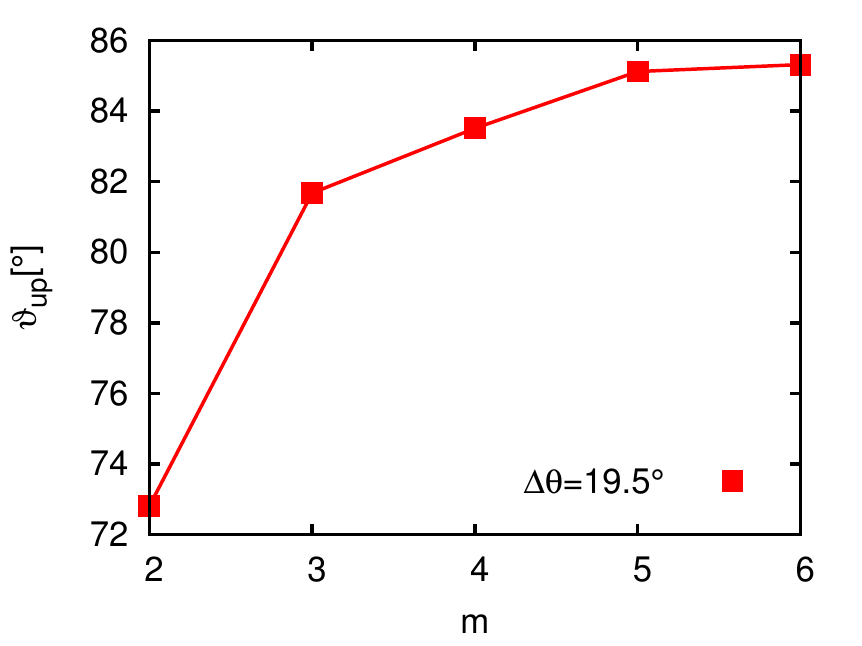}}
  \caption{
(a) 
The final value of the angle $\pafinup$ after the adsorption of the particle to the fluid
interface in the tilted state is shown for a constant aspect ratio $\Par$ and
as a function of the Janus parameter $\psjp$.
(b) 
The final angle $\pafinup$ of the particle after the adsorption to a fluid interface in the tilted state is shown for a constant $\psjp$ and depending on $\Par$.
  \label{dendpunktmkw}
  \label{pa_up_devI}
}
\end{figure}
It is shown for $\Par=1.5$ in Fig.~\ref{dendpunktcolvglmuprOri} and for several
values of $\Par\ge2$ in Fig.~\ref{dendpunktcolvglm}.  As illustrated in
Fig.~\ref{dendpunktcolvglmuprOri}, the values of the Janus parameter $\psjp$
can be divided into three ranges.  The ranges are confined by dashed, vertical
lines.  In the left range, the tilted orientation is the stable point.  In the
centered range, it cannot be concluded from the simulation results directly, if
the tilted orientation or the upright orientation minimizes the free energy.
In the right range, the upright orientation is the only final orientation found
from the simulation.  The left and the centered range
can be distinguished by using the simulation results: the adsorption trajectory
of interest is the one with $\paeffO=(10^{-4})^{\circ}$ (for the initial
orientations see \eqnref{paeffO} and the accompanying text).  If the adsorption
trajectory for this initial orientation ends up in a tilted orientation, it is
assumed that this orientation represents the global minimum of the free energy.
If this trajectory ends up in the upright orientation, then it is not possible
to see directly from the simulation results which of the two orientations
minimizes the free energy.  The horizontal line through the phase diagram
corresponds to a constant $\psjp$ and is shown in Fig.~\ref{dendpunktmdep}.  In
the simulation, the Janus parameter is restricted to the range of
$0\le\psjp\le24.4^{\circ}$.

At first, we consider the upright orientation where the Janus effect dominates.
Fig.~\ref{dendpunktcolvglmuprOri} shows that the upright state can be
reproduced for $\Par=1.5$ and $\psjp\ge15.6^{\circ}$.  This is indicated by the
second black line in Fig.~\ref{dendpunktcolvglmuprOri}.  In this range, the
upright orientation is the only orientation which is found, and thus, it can be
concluded that this is the equilibrium orientation.  The simulation results can
reproduce the upright orientation only for $\Par=1.5$ exactly.

The following step is to reproduce the parallel orientation $\pa=90^{\circ}$.
An orientation close to $90^{\circ}$ is shown in the left range of
Fig.~\ref{dendpunktcolvglmuprOri} as well as in both plots in
Fig.~\ref{pa_up_devI}.  The lines in Fig.~\ref{dendpunktcolvglmuprOri} and
Fig.~\ref{dendpunktcolvglm} show directly that the parallel orientation is
exactly reached in the limit of $\psjp=0$.  Instead of the parallel
orientation, there is a range of tilted orientations which contains the
parallel orientation in some limits.  It can be assumed from
Fig.~\ref{dendpunktmdep} by extrapolation that the parallel orientation is
reached in the limit of $\Par\to\infty$.

%
As shown in Fig.~\ref{theo1predSJ}, the transition point between both
orientations is predicted by the ``simplified Janus model'' for $\Par=1.5$ as
$\psjpt\approx10.7^{\circ}$.  This transition point cannot be explicitly
reproduced due to the extension of the centered range in
Fig.~\ref{dendpunktcolvglmuprOri}, i.e. the value of $10.7^{\circ}$ lies within
the centered range. For $\Par>1.5$ the range of area of a minimum with the
upright orientation cannot be seen.  Therefore, it is not possible to reproduce
a separation line there.

\section{Extended free energy model}
\label{sec:FE1JpFull}
The simplified free energy model described above is not sufficient to predict
the final states of the particle at the interface.  Thus, we extend it to allow
for a free rotation of the particle and take into account the interface
deformation. As it is impossible to calculate all areas of the fluid interface
analytically, we use a Monte Carlo integration of those areas from
\eqnref{FJol} which are not solvable analytically. The results of the
simulation for the particle configuration and the shape of the interface are
used to obtain their values.

The free energy given in \eqnref{FJol} and \eqnref{FEJI} is rewritten as
\begin{eqnarray}
  \label{eq:DFJ}
  \Delta\FE(\dip,\pa)&=&\FE-\FEI  \\ \nonumber &=&-\ST_{12}\iaex+\sum_{i=A,P}\ia_{1i}(\ST_{1p}-\ST_{2p})\\
      \label{eq:DFJz2}&=&\ST_{12}\left(\left(\sum_{i=A,P}\ia_{1i}\cos(\pca_i)\right)-\iaex\right)\\
      \label{eq:DFJz3}&=&\ST_{12}(\sin(\psjp)(\iaunpr-\iapr)-\iaex)
  \mbox{.}
\end{eqnarray}
$\iapr$ and $\iaunpr$ are the interfacial areas between each wetting region on
the one side and its preferred and unpreferred fluid on the other side,
respectively.  $\iaex$ is the interfacial area excluded by the adsorbed
particle, which is split into a contribution from a flat interface $\iaexflat$
and a contribution taking account to the difference caused by the interface
deformation $\Diadeltaif$. Then, we obtain
\begin{equation}
  \label{eq:DFJsplit}
\begin{split}
  \Delta\FE(\dip,\pa)=\ST_{12}(\sin(\psjp)(\iaunpr-\iapr)-\iaexflat+\Diadeltaif)
  \mbox{.}
\end{split}
\end{equation}
In order to be able to compare these interfacial areas, they are
considered in units of $\prp^2$.  The interfacial areas between a given wetting
area of the particle surface and a given fluid fulfill the relation
\begin{equation}
  \label{eq:ESJsum}
\sum_{i=A,P}\ia_{1i}=\frac{1}{2}\iaSell
=\iapr+\iaunpr
  \mbox{.}
\end{equation}
$\iaSell$ is the area of the total surface of an ellipsoid and given for
$\Par>1$ as
\begin{equation}
\begin{split}
  \label{eq:ESJ}
  \iaSell&=4\pi\pro^2\left(\frac{1}{2}+\frac{\Par^2}{4\Ecc}\ln\left(\frac{1+\Ecc}{1-\Ecc}\right)\right), \\
            &=\pi\frac{\prp^2}{\Par^2}\left(2+\frac{\Par^2}{\Ecc}\ln\left(\frac{1+\Ecc}{1-\Ecc}\right)\right),\\
            &=\pi\prp^2\left(\frac{2}{\Par^2}+\frac{1}{\Ecc}\ln\left(\frac{1+\Ecc}{1-\Ecc}\right)\right)\mbox{.}\\
\end{split}
\end{equation}
$\Ecc$ is the eccentricity of the ellipsoid, defined as
\begin{equation}
\label{eq:EE}
\Ecc=
\begin{cases}
\sqrt{1-\Par^2}, & \text{for }\Par\le1\\
\sqrt{1-\Par^{-2}}, & \text{for }\Par\ge1
\mbox{.}
\end{cases}
\end{equation}
\eqnref{DFJsplit} can be simplified with \eqnref{ESJsum} to
\begin{equation}
  \label{eq:DFJsplitf}
\begin{split}
  \Delta\FE(\dip,\pa)=\ST_{12}\left(\iaSell\left(\frac{1}{2}-\iafell\right)\cos(\pca_i)-\iaexflat+\Diadeltaif\right)
  \mbox{,}
\end{split}
\end{equation}
with the fraction of the preferred interfacial area $\iafell=2\iapr/\iaSell$.
This reduces the problem to the calculation of the three interfacial
areas
$\iafell$, $\iaexflat$ and $\Diadeltaif$. 
%
$\iaexflat$ is the excluded area of the flat fluid interface due to the
adsorbed particle. It has the shape of an ellipse with the radii $\pro$ and
$\prpeffflat$ and is given for $\dip=0$ as
\begin{equation}
  \label{eq:iaexflat}
\iaexflat=\pi\pro\prpeffflat
  \mbox{,}
\end{equation}
with
\begin{equation}
  \label{eq:prpeff}
  \prpeffflat=\frac{\prp}{\sqrt{\sin^2(\pa)+\Par^2\cos^2(\pa)}}
  \mbox{.}
\end{equation}
\eqanref{iaexflat} can be rewritten with \eqnref{prpeff} in
order to be in units of $\prp^2$,
\begin{equation}
  \label{eq:iaexflat2}
  \iaexflat=\frac{\pi\prp^2}{\Par\sqrt{\sin^2(\pa)+\Par^2\cos^2(\pa)}}
  \mbox{.}
\end{equation}

The final contributing interfacial area $\Diadeltaif$ depends on the interface
deformation indicated by $\iu$.
$\iu$ is the local level of the interface compared to the undeformed flat interface.
The general shape of the interface deformation is given by ~\cite{Stamou2000a,Lehle2008a,Loudet2005a}
	\begin{equation}
	\iu(\spv,\Aap)=
		\iuA_0\ln\frac{\spv}{\spv_0(\Aap)} 
		+\sum_{\mpo=1}^{\infty}\left(\frac{\spv_0(\Aap)}{\spv}\right)^{\mpo}(\iuA_{\mpo}\cos(\mpo(\Aap-\Aapr)) 
		+\iuB_{\mpo}\sin(\mpo(\Aap-\Aapr)))
	\mbox{.}
	\label{eq:MPD}
	\end{equation}
$\iuA_{\mpo}$ and $\iuB_{\mpo}$ are the amplitudes of the contributions in the multipol expansion.
{$\spv_0(\Aap)$ is the contact radius and depends generally on $\Aap$ for a non
spherical particle.} $\Aapr$ is the phase shift of the angle $\Aap$.  $\mpo$ is
the order of the multipole. The monopole ($\mpo=0$) is only important if the
gravitational force acting on the particle is not negligible and in this manner
leads to an interface deformation~\cite{Bleibel2013c,Oettel2008a}. In most
cases, the leading term is the quadrupole term ($\mpo=2$)~\cite{Kumar2013a}.

The final state of a symmetric Janus particle at a flat fluid interface is
found to have a leading order of a hexapolar symmetry of the interface
deformation around an ellipsoidal particle in a tilted
orientation~\cite{Rezvantalab2013b}. In this case, \eqnref{MPD} with $\mpo=3$
and $\iuB_{3}=\Aapr(\mpo=3)=0$ reduces to
\begin{equation}
\iu(\spv,\Aap)=\left(\frac{\spvO(\Aap)}{\spv}\right)^{3}\iuA_{3}\cos(3\Aap)
\mbox{ for }\radint\ge\spvO(\Aap)
  \mbox{.}
  \label{eq:MPD_mpo3}
\end{equation}
$\spvO(\Aap)$ is the contact line, where the fluid interface touches the
particle surface.  For a flat interface, the following relation can be obtained
from a geometrical consideration as
\begin{equation}
  \label{eq:spvOflat}
\spvO(\Aap)=\frac{\prpeffflat}{\sqrt{\sin^2(\Aap)\Parzd+\cos^2(\Aap)}}
  \mbox{.}
\end{equation}
$\Parzd=\prpeffflat/\pro$ is the aspect ratio of the ellipse created by the
three phase contact line of the particle surface and the fluid-fluid interface
and $\prpeffflat$ and $\pro$ are the two radii of this ellipse. It is assumed
that \eqnref{spvOflat} is approximately fulfilled for particle configurations
with an interface deformation.
%
According to \eqnref{MPD_mpo3}, $\IFDEF$ depends on the angle $\Aap$, and thus,
$\nabla\IFDEF$ is given as
\begin{equation}
\begin{split}
  \label{eq:ifdef05toJP}
\nabla\IFDEF=&\unitr\frac{\partial}{\partial\radint}\IFDEF+\unitAap\frac{1}{\radint}\frac{\partial}{\partial\Aap}\IFDEF
	\\
= &
	-\frac{3\spvO^3}{\radint^4}(\iuA_{3}\cos(3\Aap))\unitr
	-\frac{\iuA_{3}\prpeff^3(1+(1+\Parzd^2)\cos(2\Aap))\sin(\Aap)}{\radint^4(\cos^2(\Aap)+\Parzd^2\sin^2(\Aap))^{2.5}}\unitAap
  \mbox{.}
\end{split}
\end{equation}
With the relation for the orthogonal unit vectors $\unitr\cdot\unitAap=0$ and \eqnref{spvOflat} for
$\spvO$, the square of $\nabla\IFDEF$ from \eqnref{ifdef05toJP} is given as
\begin{eqnarray}
(\nabla\IFDEF)^2&=&\frac{9\spvO^6}{\radint^8}(\iuA_{3}\cos(3\Aap))^2\nonumber
+\frac{\iuA_{3}^2\prpeffflat^6(1+(1+\Parzd^2)\cos(2\Aap))^2\sin^2(\Aap)}{\radint^8(\cos^2(\Aap)+\Parzd^2\sin^2(\Aap))^{5}}\nonumber \\
&=&\frac{\iuA_{3}^2\prpeffflat^6}{\radint^8(\sin^2(\Aap)\Parzd^2+\cos^2(\Aap))^3}\nonumber 
	\left(9(\cos(3\Aap))^2+
\frac{(1+(1+\Parzd^2)\cos(2\Aap))^2\sin^2(\Aap)}{(\cos^2(\Aap)+\Parzd^2\sin^2(\Aap))^{2}}\right)\nonumber \\
\label{eq:ifdef05toJPa}
&=&\gsqifdefvor\gsqifdefr\gsqifdefphip
  \mbox{.}
\end{eqnarray}
$\gsqifdefvor\equiv\iuA_{3}^2\prpeff^6$ is a prefactor.
$\gsqifdefr\equiv\radint^{-8}$ and $\gsqifdefphip\equiv(\nabla\IFDEF)^2/(\gsqifdefvor\gsqifdefr)$
take into account the dependence of $\radint$ and $\Aap$, respectively.
With \eqnref{ifdef05toJPa}, we obtain
\begin{eqnarray}
  \label{eq:ifdef06toJPz1}
\Diadeltaif
&=&\frac{1}{2}\int_{\radint=\proeffrealif(\Aap)}^{\radintcuttoff(\Aap)}\int_{\Aap=0}^{2\pi}(\nabla\IFDEF)^2\radint
d\Aap d\radint\\
  \label{eq:ifdef06toJPz2}
&=&\frac{\gsqifdefvor}{2}\int_{\radint=\proeffrealif(\Aap)}^{\radintcuttoff(\Aap)}\int_{\Aap=0}^{2\pi}\gsqifdefr\gsqifdefphip\radint 
  d\Aap d\radint\\
  \label{eq:ifdef06toJPz3}
&=&\frac{\gsqifdefvor}{2}\int_{\radint=\proeffrealif(\Aap)}^{\radintcuttoff(\Aap)}\int_{\Aap=0}^{2\pi}\radint^{-7}\gsqifdefphip
  d\Aap d\radint\\
  \label{eq:ifdef06toJPz4}
&=&\frac{\gsqifdefvor}{12}\left(\frac{1}{\prpeff^6}-\frac{1}{\radintcuttoffO^6}\right)\int_{\Aap=0}^{2\pi}\gsqifdefphipt d\Aap\\
  \label{eq:ifdef06toJPz5}
&=&\frac{\pi\gsqifdefvor}{12}\left(\frac{1}{\prpeff^6}-\frac{1}{\radintcuttoffO^6}\right)\gsqifdefintphip(\Parzd)\\
  \label{eq:ifdef06toJP}
&=&\frac{\pi\iuA_{3}^2}{12}\left(1-\frac{\prpeff^6}{\radintcuttoffO^6}\right)\gsqifdefintphip(\Parzd)
  \mbox{.}
\end{eqnarray}
{The integration over $\radint^{-7}$ runs from the interface-particle contact
line $\proeffrealif(\Aap)$ to the cut-off radius $\radintcuttoff(\Aap)$.
The $\radint^{-7}$  dependence 
is a consequence of
{the $\radint^{-8}$ dependence from $\gsqifdefr$ and
the $\radint$ from the integration over the polar coordinates.}
The integration over $\radint^{-7}$ in \eqanref{ifdef06toJPz3} runs from the interface-particle contact
line $\proeffrealif(\Aap)$ to the cut-off radius $\radintcuttoff(\Aap)$.
The $\radint^{-7}$  dependence 
is a consequence of
the $\radint^{-8}$ dependence from $\gsqifdefr$ and
the $\radint$ from the integration over the polar coordinates.
\eqanref{ifdef05toJPa} is used in \eqnref{ifdef06toJPz2}.
\eqanref{ifdef06toJPz4} makes use of the separation between the spacial and the angular
dependence, \eqnref{spvOflat} and the relation of the cutoff radius
\mbox{$\radintcuttoff=\radintcuttoffO/\sqrt{\sin^2(\Aap)\Parzd^2+\cos^2(\Aap)}$}
  with the parameter $\radintcuttoffO$.
$\gsqifdefphipt$ is defined as
\mbox{$\gsqifdefphipt=\gsqifdefphip(\sin^2(\Aap)\Parzd^2+\cos^2(\Aap))^3$}.
\begin{equation}
  \label{eq:gsqifdintphip}
\gsqifdefintphip(\Parzd)=\pi(9+\gsqifdefnum(\Parzd))
\mbox{,}
\end{equation}
used in \eqnref{ifdef06toJPz5}, is the result of the integration over $\Aap$.
$\gsqifdefnum$ is the result of a numerical integration. $\iuA_{3}$ can be
obtained from simulations.

After the derivation of the necessary equations, we define dimensionless
quantities. A dimensionless expression for the free energy is defined as
\mbox{$\dFEG={\Delta\FE}/{(\pi\stUD\prp^2)}$}. The interfacial areas and the
parameters of the interface deformation are made dimensionless as
$\iatilde_{i}=\ia_{i}/\prp^2$ and $\iuA_{3}=\tilde{\iuA}_{3}/\prp$,
respectively.
Using these definitions, \eqnref{DFJsplitf} changes to
\begin{equation}
  \label{eq:DFJG}
  \dFEG=\frac{\Delta\FE}{\stUD\prp^2\pi} 
       =\frac{1}{\pi}\left(\left(\frac{1}{2}-\iafell\right)\iatildeSell\sin(\psjp)-\iatildeexflat+\Delta\iatilde_{\rm ifdef}\right)
  \mbox{.}
\end{equation}
$\radintcuttoffO$ is assumed to be $\infty$.
The interfacial areas in \eqnref{DFJG} can be substituted with \eqnref{ESJ}, \eqnref{iaexflat2} and \eqnref{ifdef06toJP}. Using the relations \eqnref{EE} and \eqnref{gsqifdintphip}, \eqnref{DFJG} can be rewritten as
\begin{equation}
\begin{split}
  \label{eq:DFJG2}
  \dFEG=&\left(\frac{1}{2}-\iafell\right)\sin(\psjp) \\ &\left(\frac{2}{\Par^2}+\frac{1}{\sqrt{1-\Par^{-2}}}\ln\left(\frac{1+\sqrt{1-\Par^{-2}}}{1-\sqrt{1-\Par^{-2}}}\right)\right)
	-\frac{1}{\Par\sqrt{\sin^2(\pa)+\Par^2\cos^2(\pa)}}+\frac{\iuAtilde_{3}^2}{12}\gsqifdefintphip(\Parzd)\\
       =&\left(\frac{1}{2}-\iafell\right)\sin(\psjp) 
	\left(\frac{2}{\Par^2}+\frac{1}{\sqrt{1-\Par^{-2}}}\ln\left(\frac{1+\sqrt{1-\Par^{-2}}}{1-\sqrt{1-\Par^{-2}}}\right)\right)\\
        &-\frac{1}{\Par\sqrt{\sin^2(\pa)+\Par^2\cos^2(\pa)}}+\frac{\iuAtilde_{3}^2}{12}(9+\gsqifdefnum(\Parzd))
  \mbox{.}
\end{split}
\end{equation}
For the upright orientation, the free energy can be calculated exactly as shown
in \eqnref{FadN}.  For the tilted orientation, the factor $\iafell$ cannot be
calculated analytically. It is calculated with the Monte Carlo method. The
factor $\iuAtilde_{3}$ is read out from the interface deformation obtained by
the simulation.
The result of \eqnref{DFJG2} for the tilted orientation is compared with the
one for the upright orientation in order to find the global minimum of the free
energy.

\begin{figure}
\centering
\subfigure[]
	{\label{FEm2}
	\includegraphics[width=7cm]{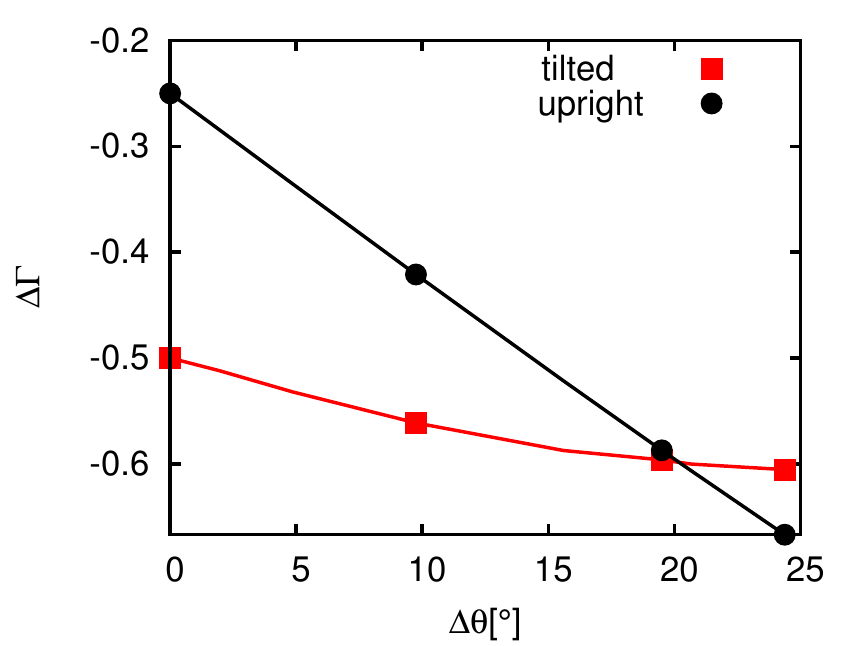}}
\qquad
\subfigure[]
        {\label{FESimuPhase}
        \includegraphics[width=7cm]{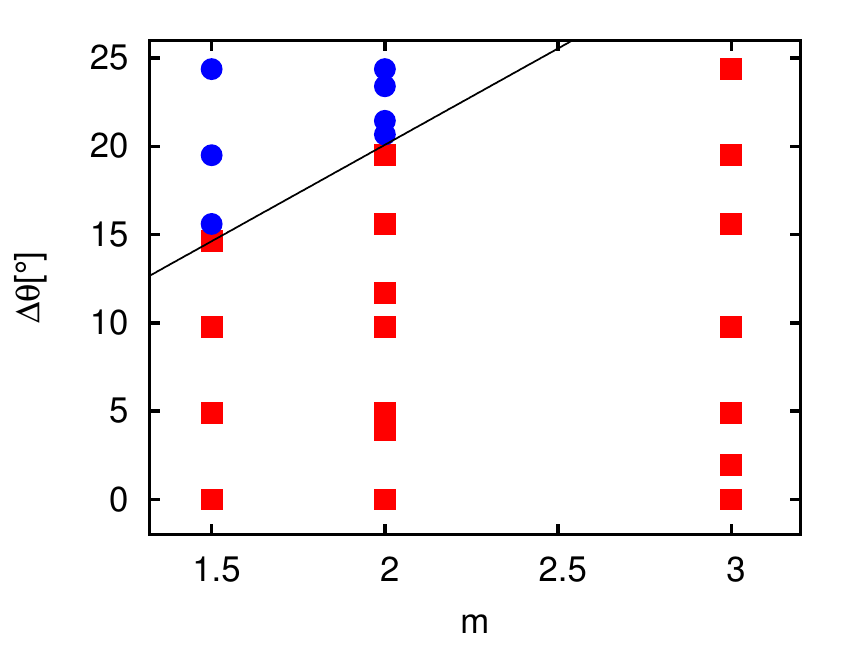}}
	\caption{(a) The dimensionless free energy $\dFEG$ (see \eqnref{DFJG}) is shown for
	the tilted state as well for the upright state for $\Par=2$.  A
	transition point is found at $\psjpt\approx20^{\circ}$. For
	$\psjp<\psjpt$ the tilted state minimizes the free energy. The upright
	state minimizes the free energy above $\psjpt$.\\
        (b) The regions in the $\Par$-$\psjp$ phase diagram where the tilted and
	upright state minimize the free energy are shown by squares and
	circles, respectively.  The black line separates both areas.  }
\end{figure}

Figure~\ref{FEm2} shows the dimensionless free energy $\dFEG$ as a function of
$\psjp$ for $\Par=2$.  It compares the values of $\dFEG$ for the tilted state
and the upright state.  A transition point is found at
$\psjpt\approx20^{\circ}$. For $\psjp<\psjpt$ the geometry dominates and the
upright orientation minimizes the free energy. For larger values of the Janus
parameter, the wettability effect dominates and the upright orientation poses
the free energy minimum.
The phase diagram depending on $\psjp$ and $\Par$ is shown in
Fig.~\ref{FESimuPhase}. The results are obtained from the simulation of a Janus
particle adsorption to a flat interface and \eqnref{DFJG2}. The study is done
in the ranges of $0\le\psjp\le25^{\circ}$ and $1.5\le\Par\le6$.  The regions
where the shape dominates and is in the tilted state are shown by red squares. The
Janus dominated regions with the upright orientation as the free energy minimum
are shown by blue circles. For $\Par\ge3$ the tilted state is found in the
whole simulated region. For smaller aspect ratios transition points are found.
The transition value of $\psjp$ increases with an increasing aspect ratio
$\Par$.

\section{Adsorption trajectories}
\label{sec:adsorption}
Finally, we study the trajectories of the adsorption of a single Janus particle
to a flat fluid interface.  We present three different situations where the
interplay between particle shape and wettability difference has a different
impact.  The three examples differ in the choice of the two parameters $\Par$
and $\psjp$.

The first situation with the parameters $\Par=2$ and $\psjp\approx2^\circ$
is presented {in} Fig.~\ref{adsorb_a01_a05_02}.
The dashed and dotted
lines show the adsorption trajectories for the PIO (``preferred initial
orientation'', see Fig.~\ref{fig:geo}) and the UIO (``unpreferred initial orientation''),
respectively. The square indicates the ending point of both
dashed and dotted lines.
The circle and blue triangles show the exclusive
end points of dashed and dotted lines, respectively.
The solid lines indicate the points where the particle touches the
interface. This is the initial condition for the PIO and the UIO as shown in Fig.~\ref{fig:geo}.
The Roman numerals relate the points for the final states in
Fig.~\ref{adsorb_a01_a05_02} 
to the sketches in Fig.~\ref{stablepat}.
\begin{figure}
  \centering
  \subfigure[]
  {\label{adsorb_a01_a05_02}\includegraphics[width=7cm]{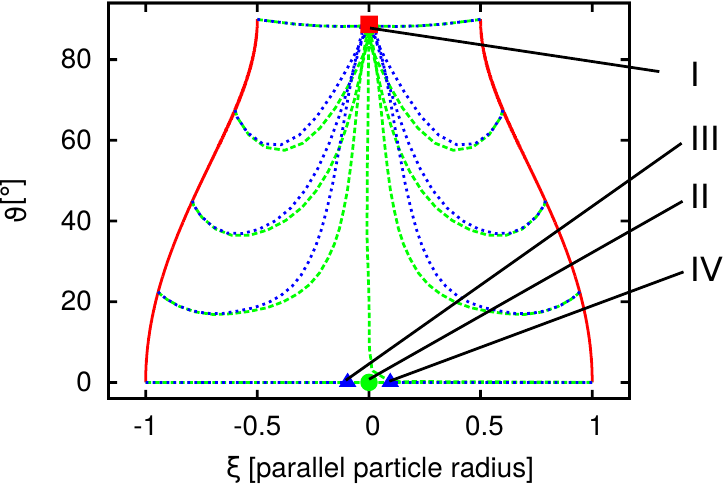}}
\qquad
  \subfigure[]
  {\label{stablepat}\includegraphics[width=5.4cm]{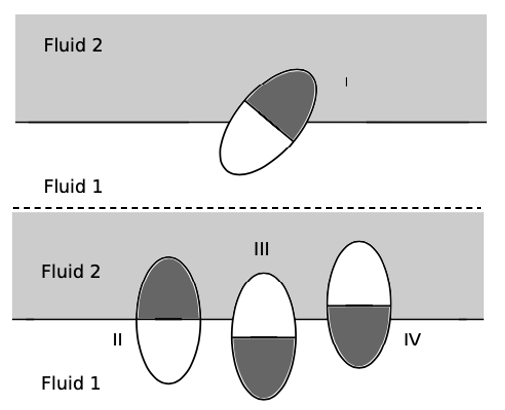}}

  \caption{
	  (a) Adsorption trajectories ($\paeff$-$\dip$-plot) of a single Janus particle at a flat interface for $\Par=2$ and $\psjp\approx2^\circ$.
The dashed and dotted lines represent the trajectories for the PIO and UIO,
respectively. The following types of end points or final states (characterized by $\dip$ and $\paeff$) are shown: The square indicates the end point of both
dashed and dotted lines. The circle and triangles show the exclusive
end points of dashed and dotted lines, respectively.
The Roman numerals relate the points for the final states to the following sketches:
	(b) Sketch of the tilted state (``I'', $0<\paeff<90$ {and $\dip=0$}) corresponding to the square in (a) (upper end point) and
the metastable states for different cases of the upright
orientation $\paeff=0$ (``II'').
Left particle: every wetting area is immersed completely in its preferred fluid.
	$\pa=0$, $\dip=0$ (related to the circle in (a)).
Middle (``III'') and right (``IV'') particle: one wetting area is completely and the other one
partly inside its unpreferred fluid.
$\pa=180^\circ$, $\pm\dipfinunpr$ (related to both triangles
	in (a) and given in \eqnref{dipfinunpr}).
The interface is flat for all three final states.
}
  \label{FirstExample1pJanus_3casesI}
\end{figure}
All trajectories with $\paeffO>0$ end up in a tilted state (as shown by the
square in Fig.~\ref{adsorb_a01_a05_02} and state ``I'' in Fig.~\ref{stablepat})
characterized by the final position of $\dipfinalup=0$.  The final orientation
is tilted and close to the parallel orientation found for a particle with a
uniformly wetting surface. This is close to the state with the largest
interfacial area occupation.  The latter was achieved in the case for a
uniformly wetting particle, where the particle is oriented parallel to the
interface. Even the line of $\paO=(10^{-4})^{\circ}$ ends up in state ``I''
suggesting the existence of a free energy maximum for Janus particles. This is in agreement with Fig.~\ref{FESimuPhase}.

For the trajectories of the initial orientation of $\paeffO=0$, the particle
does not rotate, but stays in the upright orientation.  One needs to
distinguish between the PIO ($\paO=0$) and the UIO ($\paO=180^{\circ}$).  In
the PIO, the particle ends at the point $\paeffO=0$ and
$\dipfinalpr=0$.
This state is indicated by
the circle in Fig.~\ref{adsorb_a01_a05_02} and the left particle (state ``II'')
in Fig.~\ref{stablepat}. Each wetting area is completely immersed in its
preferred fluid {and the border between both wetting areas is exactly in
contact with the interface}.  The interface is undeformed in this state.  In
the UIO, the final position splits as $\pm\dipfinunpr$ for the two adsorption
trajectories coming from  both sides of the interface.  $\dipfinunpr$ is the
absolute value of the position for the upright orientation in the UIO.
For a positive sign of $\dipO$, the particle ends up in a distance of
$\dipfinunpr$ above the interface and for a negative sign of $\dipO$ in a
distance of $\dipfinunpr$ below the interface.  This state is indicated by the
two triangles in Fig.~\ref{adsorb_a01_a05_02} for both values of $\dipfinunpr$
and the centered and right particle in Fig.~\ref{stablepat}.  A relation
for $\dipfinunpr$ can be obtained from the geometrical consideration of an
ellipsoidal particle as
\begin{equation}
  \label{eq:dipfinunpr}
  \dipfinunpr=\frac{\Par\prp\tan(\psjp)}{\sqrt{1+\Par^2\tan^2(\psjp)}}
  \mbox{.}
\end{equation}
In this configuration ($\paO=180^{\circ}$ and $\pm\dipfinunpr$), the Young
equation can be fulfilled by a flat fluid interface.
In the recent example with $\Par=2$, $\psjp=1.95^\circ$ and $\prp=1$
  (the latter one due to the
  fact that $\dipfinunpr$ is like $\dip$ in units of $\prp$), we obtain
$\dipfinunpr=6.8\cdot10^{-2}$ from \eqnref{dipfinunpr}. This is close to the value from the simulation
$\dipfinunpr\approx9.5\cdot10^{-2}$
(the blue triangle in Fig.~\ref{adsorb_a01_a05_02}). 
The deviation is a result of the discretization of the particle on the lattice.

\begin{figure}
  \centering
  \subfigure[]
  {\label{adsorb_a01_a02_01}\includegraphics[width=7cm]{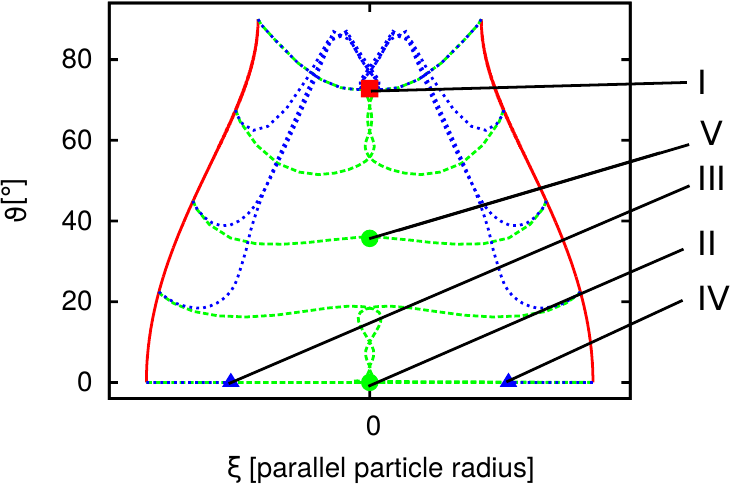}}
\qquad
  \subfigure[]
  {\label{adsorb_a09_a02_01}\includegraphics[width=7cm]{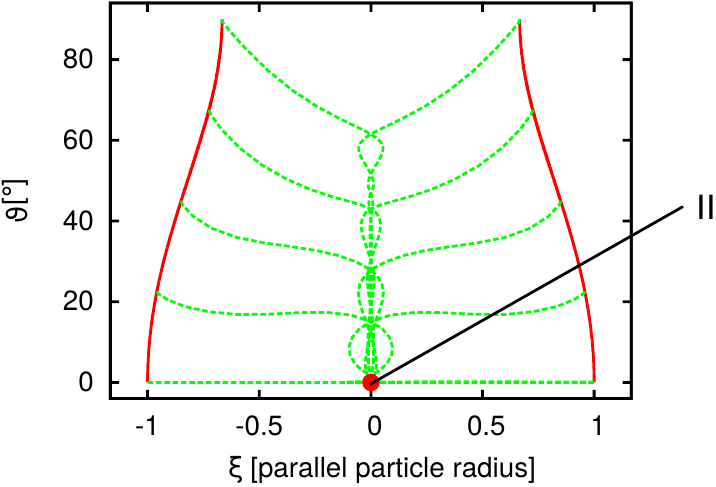}}
  \caption{
$\paeff$-$\xi$-plots for the adsorption of a single Janus with different values of $\Par$ and $\psjp$.
Line and end point colors are identical to Fig.~\ref{adsorb_a01_a05_02}.  The
	Roman numerals relate the points for the final states in
	Fig.~\ref{stablepat} and Fig.~\ref{stablepat}.  State ``V'' is a
	tilted state similar to state ``I'', but with a different angle $\pa$.
(a) ($\psjp\approx19.5^\circ$, $\Par=2$) The structure of the $\pa$-$\dip$ plot
	is more complicated as compared to Fig.~\ref{adsorb_a01_a05_02}.  For
	the PIO, there are two different tilted states. Furthermore, the range
	of initial orientations for the final upright orientation is extended.
(b) ($\Par=1.5$ and $\psjp\approx19.5^\circ$) Only one final point is there for
	the upright position $\pafin=\pafinupr=0$ as shown by the left particle
	in Fig.~\ref{stablepat}.
}
  \label{FirstExample1pJanus_3casesII}
\end{figure}

Figure~\ref{adsorb_a01_a02_01} shows the adsorption trajectories for $\Par=2$
and $\psjp\approx19.5^\circ$.  The Janus effect is stronger as compared to the
previous example.  Furthermore, the structure of end points is more complex.
Besides the upper point $\pafinup$ (state ``I'') and the upright point
$\pafinupr$ (states ``II'', ``III'' and ``IV''), there is an additional end
point related to the orientation $\pafinmiddle$ (``second tilted orientation'',
state ``V'').
It is a tilted orientation similar to the one of state ``I'', but with a
different value of the angle.
Its value is located between the two other
  orientations as $\pafinupr<\pafinmiddle<\pafinup$.
The final point of a given adsorption trajectory depends on
the initial orientation as follows:
for the PIO, the final position is given as $\dip=0$ for all initial orientations.
The trajectories for the two upper starting angles ($\paO\ge67.5^{\circ}$) end up in the upper final point (state ``I'').
The trajectory of $\paO=45^{\circ}$ ends up (as the only one from the considered
initial orientations defined in \eqnref{paeffO}) in the ``second tilted orientation''.
The trajectories for the two lower values of $\paO$ end up in the upright
orientation $\pafinupr$ (state ``II''). In contrast to the situation of
Fig.~\ref{adsorb_a01_a05_02}, there is an extended range of initial orientations
which end up finally in the upright orientation.

For the UIO, the ending point structure is completely different from the PIO.
The trajectories for the reference orientations $\paeffO\le22.5^{\circ}$ end up
in the upper final point. For an initial orientation of $\paO=180^{\circ}$,
the particle keeps its orientation during the adsorption process. The position
is predicted from \eqnref{dipfinunpr} $\dipfinunpr\approx5.78\cdot10^{-1}$ and
from the simulation results $\dipfinunpr=6.22\cdot10^{-1}$.  All UIO
trajectories of $\paO\neq180^{\circ}$ end up in $\dip=0$. Note that state ``V''
can be found in several regions in the $\Par$-$\psjp$ diagram.  It is shown in
the calculations in section~\ref{sec:FE1JpFull} that there is no region in the
$\Par$-$\psjp$ diagram , where state ``V'' minimizes the free energy.

A fully Janus dominated case is shown in Fig.~\ref{adsorb_a09_a02_01} for
$\Par=1.5$ and $\psjp\approx19.5^\circ$.  All studied trajectories end up in
the upright orientation (state ``II'').  There is no tilted state found for any
of the initial orientations.  The anisotropy is so weak that the free energy
gain due to the excluded interfacial area $\ia_{ex}$ for a final orientation of
$\pa>0$ is small. However, the Janus effect ($\psjp$) is sufficiently strong to
enforce an orientation where each wetting area is in its preferred fluid.


\section{Conclusions}
We studied the equilibrium configuration and adsorption process of ellipsoidal
Janus particles at a fluid-fluid interface.  We presented free energy models
with and without consideration of the deformation of the fluid interface and
compared the final states of the particle at the fluid interface obtained from
the lattice Boltzmann simulations with the predictions from the free energy
calculation. The model without interface deformation is only able to predict
the simulation results within a specific range, whereas the improved model
which takes into account the full rotation of the particle and the
deformability of the interface shows a good agreement with our simulation
results. 

We show that the equilibrium state of a Janus ellipsoid which is determined by the minimum
of the free energy is a tilted orientation for large aspect ratios and small
wettability differences, where the shape dominates. In case of small aspect
ratio and large wettability difference, the Janus effect dominates and the
particle is in an upright orientation. Thus, the equilibrium state of an
ellipsoidal Janus particle at a fluid interface can be tuned by the choice of
aspect ratio and wettability contrast.


Finally, we studied the adsorption trajectories of ellipsoidal Janus particles
for three representative cases of the particle aspect ratio and wettability
contrast. The interplay of these two parameters does not only influence the
final state the sytsem will achieve, but also determines the complex dynamics
of the rotation and movement of the particle. Furthermore, we demonstrated that
the adsorption trajectory of a symmetric, ellipsoidal Janus particle can show
up to three metastable points whereas the uniformly wetting counterpart is
known to have only a single metastable point. 

\vspace*{2mm}
\begin{acknowledgments}
	Financial support is acknowledged from the German Research Foundation
	(DFG) through priority program SPP2171 (grant HA 4382/11-1).
	We thank the J\"ulich
	Supercomputing Centre and the High Performance Computing Center
	Stuttgart for the technical support and allocated CPU time.
\end{acknowledgments}

\end{document}